\documentclass[aps,prl,twocolumn,showpacs,groupedaddress]{revtex4}  
\usepackage{graphicx,color,dcolumn,bm,longtable,mathptmx,float,titlesec,lipsum}
\usepackage{amsfonts,amsmath,amssymb}
\usepackage{amsfonts,amsmath,amssymb}
\begin{document}
\title {Inducing and Manipulating Magnetization in Two-Dimensional ZnO by Strain and External Gating}
\author {P. Taivansaikhan,$^{1}$ T. Tsevelmaa,$^2$ S. H. Rhim,$^{2}$ S. C. Hong,$^2$ and D. Odkhuu$^{1}$}
\email{odkhuu@inu.ac.kr}
\affiliation{
$^1$Department of Physics, Incheon National University, Incheon 22012, South Korea \\
$^2$Department of Physics, University of Ulsan and EHSRC, Ulsan 44610, South Korea }
\begin{abstract}

Two-dimensional structures that exhibit intriguing magnetic phenomena such as perpendicular magnetic anisotropy and
switchable magnetization are of great interests in spintronics research.
Herein, the density-functional theory studies reveal the critical impacts of strain and external gating on vacancy-induced
magnetism and its spin direction in a graphene-like single layer of zinc oxide (ZnO).
In contrast to the pristine and defective ZnO with an O-vacancy, the presence of a Zn-vacancy
induces significant magnetic moments to its first neighboring O and Zn atoms due to the charge deficit.
We further predict that the direction of magnetization easy axis reverses from an in-plane to perpendicular orientation under a practically achieved biaxial compressive strain of $\sim$1--2\%
or applying an electric-field by means of the charge density modulation.
This magnetization reversal is driven by the strain- and electric-field-induced changes in the spin-orbit coupled \emph{d} states of the first-neighbor Zn atom to the Zn-vacancy.
These findings open interesting prospects for exploiting strain and electric-field engineering to manipulate magnetism and magnetization orientation of two-dimensional materials.

\end{abstract}
\maketitle
\section{\textbf{INTRODUCTION}}
During the past two decades since the successful isolation of graphene from graphite \cite{nov04,nov05},
the exploration of two-dimensional (2D) layered materials
has received much attention and is likely to remain one of the leading topics in material science into foreseeable future.
Although there have been a number of experimental successes and theoretical predictions on discovering 2D materials, including graphene, \textit{hexagonal} boron nitride (\emph{h}-BN), and transition metal dichalcogenides (TMDs) \cite{nov04,nov05,nagashima95,mak10,geim13}, their potentials in spintronic applications have been limited by the lacks of spin orders and possibilities of magnetization flip.
Hence, the current research in this field primarily focuses on seeking a novel 2D material with intriguing magnetic properties \cite{leht04,han13,gon16,od16},
which would offer great advantages for exploring the internal quantum degrees of freedom of electrons and their potentials in nanoscale-down spin-based electronic applications.

Zinc oxide (ZnO), as a 2D material, can possibly be a promising candidate in the same regard,
since it has been recently reported that a (0001) oriented ultrathin ZnO film can be transformed into a new form of stable graphite-like structure \cite{tusche07,freeman06,tu06,tu10,wu11}.
One to three or four atomic layers of ZnO(0001) have been successfully synthesized on a (111) surface of Ag \cite{tusche07} and Au substrates \cite{deng13,tumino16,lee16},
where Zn and O atoms are arranged in the lateral lattice as a honeycomb shape, similar with the \emph{h}-BN structure.
Theoretical studies have also predicted that thin films of ZnO prefer a graphite-like layered structure when the thickness is less than three atomic layers \cite{tu06,tu10},
and the stability of even thicker graphitic layers can be improved by an epitaxial strain \cite{wu11}.
More remarkably, an epitaxial growth of ZnO monolayer on graphene has been very recently demonstrated in atomic resolution transmission electron microscopy (ARTEM) measurements \cite{hong17}.

Numerous reports, on the other hand, have already shown that intrinsic defects and impurities can readily give rise to magnetism in ZnO allotropies, including monolayer, graphitic layers, thin films, nanowires and nanoribbons \cite{wang08,xu08,botello08,khalid09,topsakal09,kim09,schmidt10,xing11,guo12,tan15,fang16}.
For instance, nonmetal (C, B, N, and K)-\cite{guo12,fang16} and rare-earth metal (Ce, Eu, Gd, and Dy)-doped ZnO monolayer \cite{tan15} exhibits magnetic instability via first-principles calculations.
Moreover, it has been shown by both experimentally \cite{xu08,khalid09,xing11} and theoretically \cite{wang08,topsakal09,kim09} that the Zn vacancy induces ferromagnetism in either thin films \cite{xu08,khalid09,kim09,xing11} and graphitic layers \cite{wang08,topsakal09} of ZnO without the need of magnetic impurities.

In this article, we perform first-principles calculations to reveal the synergistic effect of biaxial strain on magnetism and its spin direction of a graphene-like single layer of ZnO induced by a Zn-vacancy.
Through the analyses of the spin-orbit Hamiltonian matrix elements and atom- and orbital-decomposed magnetic anisotropy energy (MAE),
we elucidate the underlying mechanism of magnetization reversal,
which occurs at the compressive biaxial strain of only $\sim$1--2\%,
in terms of the strain-induced changes in the spin-orbit coupled \emph{d} states of the vacancy-neighbored Zn atom.
Moreover, this anisotropic magnetism is identified to even be reversible upon the change in charge carrier density,
suggesting the possibilities of chemical doping or magnetoelectric control of magnetization reorientation.

\section{\textbf{COMPUTATIONAL DETAILS}}
Density-functional theory (DFT) calculations were performed using the Vienna \emph{ab initio} simulation package (VASP) with plane-wave basis set \cite{Kresse96a,Kresse96b}. The projector augmented wave (PAW) potentials \cite{Blochl994} were used to describe the core electrons, and the generalized gradient approximation (GGA) of Perdew, Burke, and Ernzernhof (PBE) was adopted for exchange-correlation energy \cite{PBE1996}.
As illustrated in Fig. 1(a), a 4$\times$4 lateral unit cell of the \emph{hexagonal} lattice was chosen as a model geometry of a single-layer ZnO.
We also consider an intrinsic defect of an oxygen vacancy [Fig. 1(b)] as well as a zinc vacancy [Fig. 1(c)] in ZnO monolayer,
which occurs notably often during sample preparation and treatments \cite{tuomisto03,tuomisto05,janotti07,chen07,zubiaga08,dong10,khan13}.
In order to avoid spurious interactions between periodic images of ZnO layer,
a vacuum spacing of no less than 15 {\AA} perpendicular to the plane was employed.
An energy cutoff of 500 eV and a 5$\times$5$\times$1 \emph{k} mesh were imposed for the ionic relaxation, where forces acting on atoms were less than 10$^{-2}$ eV/{\AA}.
The MAE is obtained based on the total energy difference when the
magnetization directions are in the \emph{xy} plane (\emph{E}$^\|$) and along the \emph{z} axis (\emph{E}$^\bot$),
MAE $=$ \emph{E}$^\|$ -- \emph{E}$^\bot$, where convergence with respect to the \emph{k} point sampling is ensured.
To obtain reliable values of MAE, the Gaussian smearing method with a smaller smearing of 0.05 and dense \emph{k} points of 15$\times$15$\times$1 were used in noncollinear calculations, where the spin-orbit coupling (SOC) term was included using the second-variation method employing the scalar-relativistic eigenfunctions of the valence states \cite{Koelling1977}.

\begin{figure} [t]
  \centering
   \includegraphics[width=1\columnwidth]{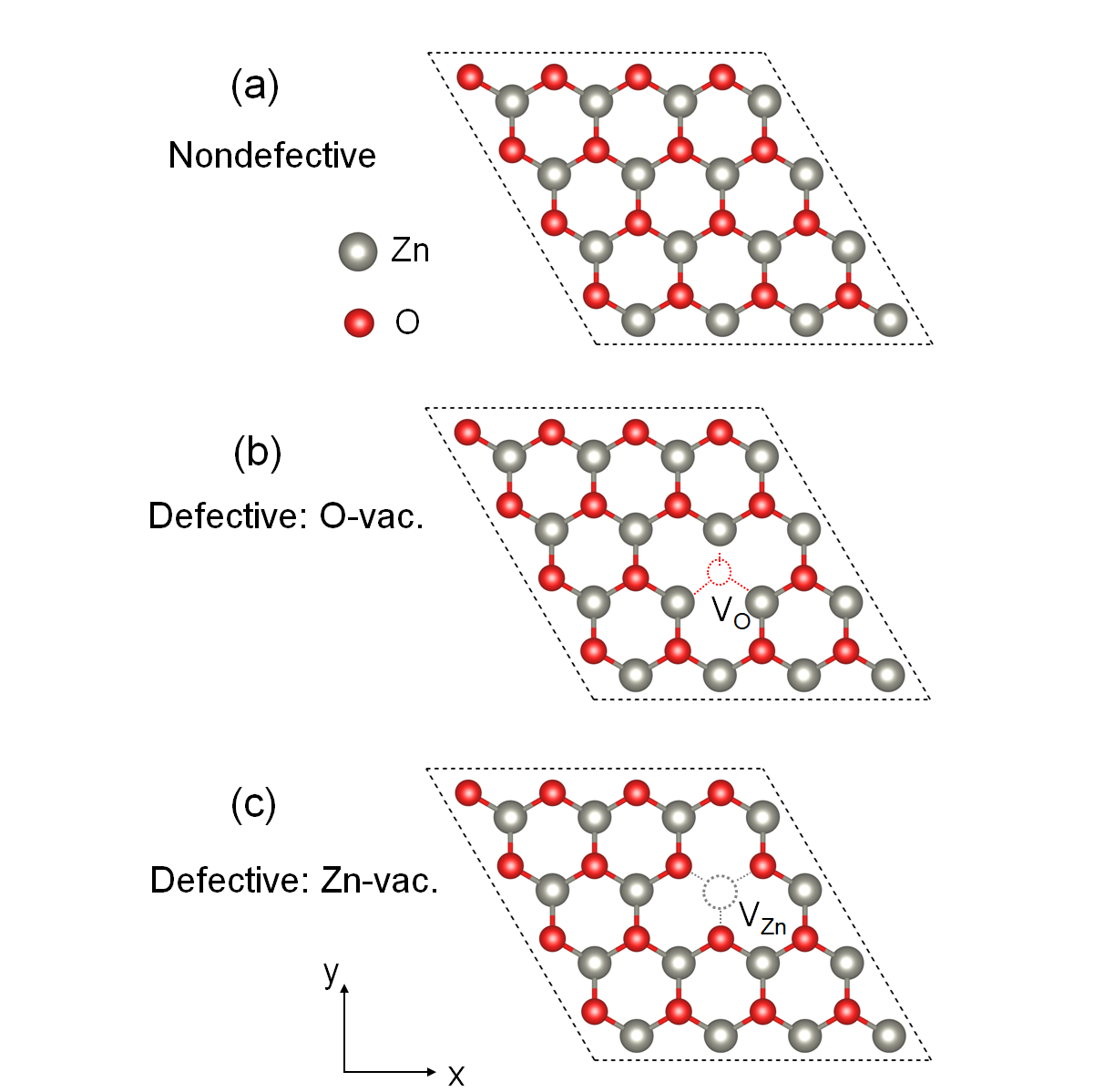}
  \caption {Top views of the atomic structures for (a) a single-layer ZnO with (b) a O-vacancy and (c) a Zn-vacancy.
  The larger gray and smaller red spheres are Zn and O atoms, respectively. In (b) and (c), the red and black dotted-circles correspond to the O and Zn vacancies, respectively.}
\end{figure}

\begin{table} [b]
\caption{Optimized in-plane lattice \emph{a} (\AA),
formation energy $H_f$ (eV),
magnetic moment $\mu$ ($\mu_\textrm{B}$) and
charge $\rho$ (\emph{e}) of Zn and O atoms
in the nondefective ($V_{\textrm{free}}$)
and defective ZnO with an O- ($V_{\textrm{O}}$) and a Zn-vacancy ($V_{\textrm{Zn}}$).}
\begin{ruledtabular}
\begin{tabular}{ccccccccccc}
 & \emph{a} & $H_f$ & $\mu^{\textrm{Zn}}$ & $\mu^{\textrm{O}}$ &$\mu^{\textrm{Total}}$&
$\rho^{\textrm{Zn}}$ & $\rho^{\textrm{O}}$ \\
\hline
$V_{\textrm{free}}$ & 3.290 & -- &0.0 & 0.0 & 0.0 & 10.81 & 7.18 \\
$V_{\textrm{O}}$ & 3.248 & --3.75 & 0.0 & 0.0 & 0.0 & 10.86 & 7.15 \\
$V_{\textrm{Zn}}$ & 3.296 & --5.43 & 0.04 & 0.32 & 2.0 & 10.79 & 6.97 \\
\end{tabular}
\end{ruledtabular}
\end{table}

\section{\textbf{RESULTS AND DISCUSSION}}
We first optimize the lattice parameters of the nondefective and defective ZnO monolayer and the results are shown in Table I.
The lattice constant of the nondefective ZnO monolayer
is 3.29 \AA, which agrees pretty well with experimental values of $\sim$3.29--3.30 {\AA} \cite{lee16,tusche07,hong17}, superior to the previous theories ($\sim$3.20--3.28 {\AA}) \cite{tu06,topsakal09,tu10}.
While the lattice constant of the nondefective ZnO remains almost unchanged in the presence of a Zn-vacancy,
it is reduced by $\sim$1\% when an O-vacancy is introduced (in a 4$\times$4 unit cell of ZnO monolayer).
The former is due to the direct repulsive interaction between O--O anions without Zn mediation.
As presented in Table I, the further calculations of formation energy $H_f$, defined as $H_f = E_{\textrm{Nondef}}-E_{\textrm{Def}}-E_{\textrm{Zn/O}}$, where $E_{\textrm{Nondef}}$, $E_{\textrm{Def}}$, and $E_{\textrm{Zn/O}}$ represent the total energies of the nondefective and defective ZnO with a Zn/O-vacancy, and a Zn/O atom in its bulk/O$_2$-molecule form,
indicate the favorable formation of a vacancy in ZnO, particularly for the Zn-site vacancy, as reported in experiments \cite{tuomisto03,tuomisto05,janotti07,chen07,zubiaga08,dong10,khan13}.

\begin{figure} [t]
  \centering
 \includegraphics[width=1\columnwidth]{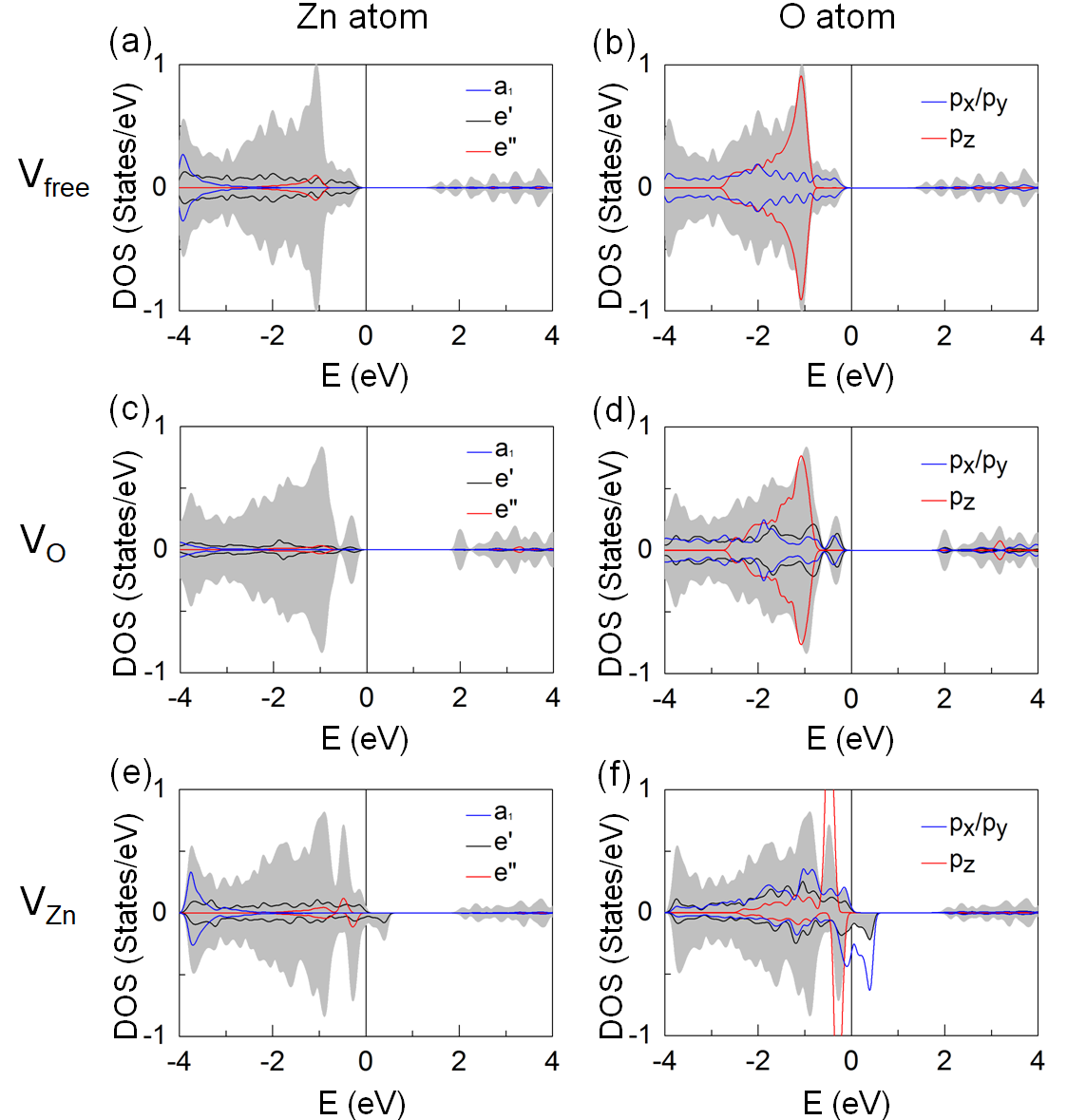}
  \caption {Spin- and orbital-decomposed PDOS of (a) Zn and (b) O atoms of the nondefective ZnO monolayer. The same PDOS for the defective ZnO with (a) and (b) a O-vacancy and (e) and (f) a Zn-vacancy.
  The singlet $a_1$ ($d_{z^2}$) and doublet $e'$ ($d_{xy/x^2−y^2}$) and $e''$ ($d_{xz/yz}$) orbital states of Zn atom are shown in blue, black, and red lines, respectively. The $p_{x,y}$ and $p_{z}$ orbital states of O atom are shown in blue and red lines, respectively.
For reference, the total DOS per atom (shaded area) is also shown in each panel. The Fermi level is set to zero in energy.}
\end{figure}

The significance of induced magnetism, as fairly addressed in experimental \cite{xu08,khalid09,xing11} and previous theoretical studies \cite{wang08,topsakal09,kim09}, is then justified.
Note that the magnetic moments of the first-neighboring Zn and O atoms to the Zn-vacancy site
are the most prominent beyond those (negligibly small) of the other sites away from the vacancy.
We thus refer the results and discussion here and hereafter mainly to those corresponding to the first-neighboring Zn and O atoms.
As seen in Table I, the nondefective and defective ZnO with an O-vacancy behave nonmagnetic nature as in bulk \cite{hong07,chen13},
but the Zn and O atoms in the Zn-vacancy defected ZnO exhibit significant spin magnetic moments of 0.04 and 0.32 $\mu_\textrm{B}$, respectively.
This vacancy-induced magnetic instability
simply reproduces the results reported in previous \emph{ab initio} calculations, where the total magnetization in an unit cell is 2 $\mu_\textrm{B}$ \cite{wang08,topsakal09,kim09}.
Moreover, our results in ZnO monolayer are more and less comparable with those ($\mu^{\textrm{Zn}}$ $=$ $\sim$0.04 and $\mu^{\textrm{O}}$ $=$ $\sim$0.14--0.48 $\mu_\textrm{B}$) in ZnO thin films \cite{wang08,kim09}.
The vacancy defect driven magnetism can be understood in terms of the spin-polarized charge excess or deficit, where the total magnetization in an unit cell is determined by the unpaired electron counts.
Our additional calculations with an injected charge show that
such a charge deficit is compensated by an extra charge of +2$e^{-}$ per Zn-vacancy and thereby the induced magnetism disappears.

\begin{figure} [t]
  \centering
 \includegraphics[width=1\columnwidth]{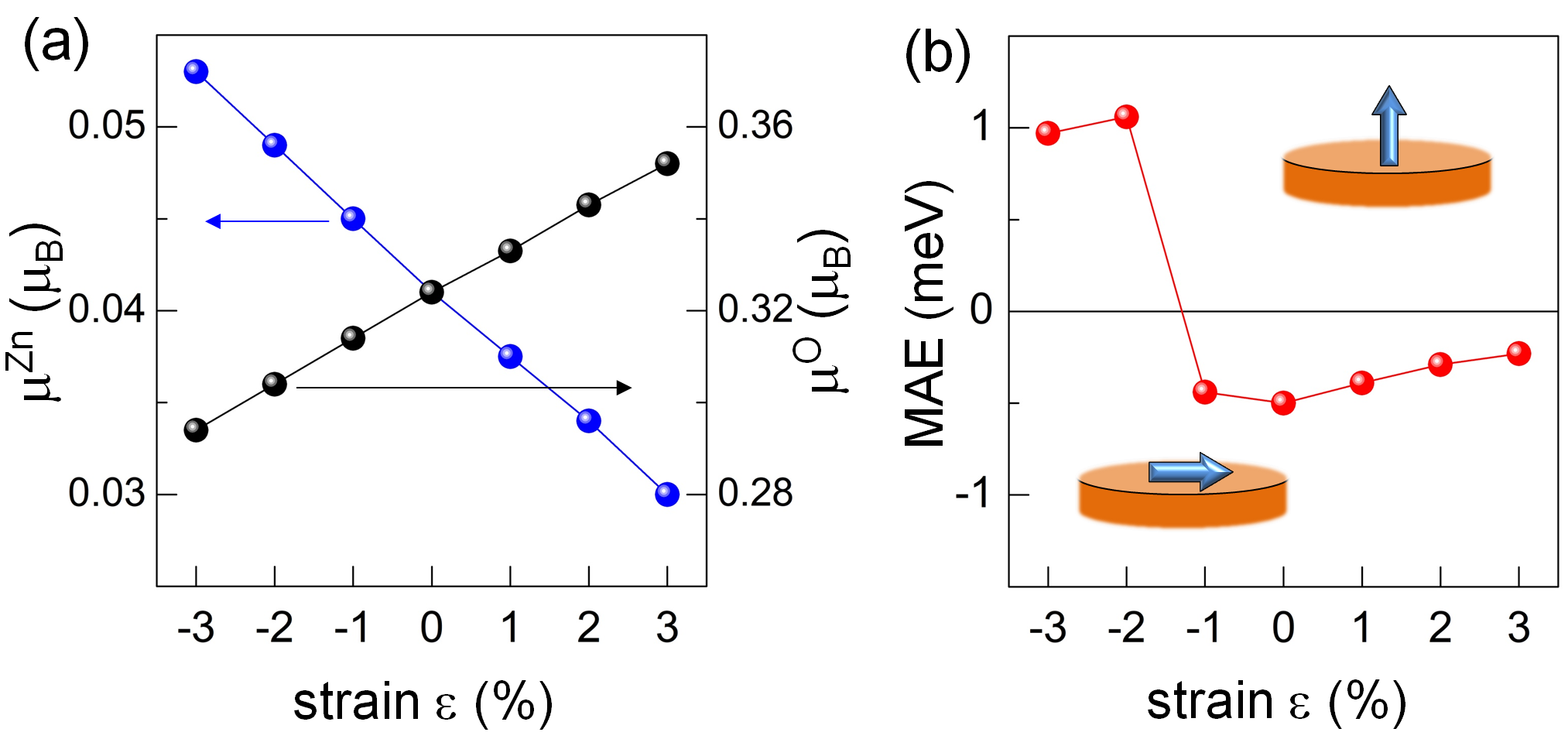}
  \centering
  \caption{(a) Magnetic moments $\mu$ of Zn (blue) and O (black) atoms and (b) MAE of a single-layer ZnO with a Zn-vacancy as a function of strain ($\varepsilon$). The negative and positive $\varepsilon$ correspond to the compressive and tensile strain, respectively.
  The insets in (b) show the schematics for magnetization switching from in-plane (bottom in negative MAE) to out-of-plane magnetization (top in positive MAE) at a compressive biaxial strain of $\sim$1--2\% .}
\end{figure}

To better understand the Zn-vacancy induced magnetism,
we show the projected density of states (PDOS) of the Zn \emph{d} and O \emph{p} orbital states in Figs. 2(e) and 2(f), respectively, in comparison to those
of the nondefective [Figs. 2(a) and 2(b)] and O-vacancy defected ZnO [Figs. 2(c) and 2(d)]. In hexagonal lattice, the five \emph{d}-orbital states split into one singlet
$a_1$ or $m=0$ ($d_{z^2}$) and two doublets $e'$ or $m=\pm2$ ($d_{xy/x^2−y^2}$) and $e''$ or $m=\pm1$ ($d_{xz/yz}$).
Both the nondefective and O-vacancy defected ZnO exhibit wide-bandgap insulating electronic features,
where the Zn \emph{d} orbital states are fully occupied while the conduction band is mainly the characteristics of the O \emph{p} states.
Furthermore, the spin subbands of the majority- and minority-spin states are essentially degenerate.
On the other hand, for the Zn-vacancy defected ZnO the majority- and minority-spin states of the Zn and O atoms reflect unequally (i.e., net exchange splitting); The minority-spin filled bands of the Zn $e'$ and O $p_{x,y}$ states move simultaneously upward across the Fermi level in energy relative to the corresponding majority-spin states,
thus exhibits a half-metallic behavior.
This is mainly due to the charge delocalization of the minority-spin electrons of the Zn and O atoms due to the broken charge quantization associated with the absence of a Zn atom,
which are most likely distributed toward the Zn-vacancy site.
From the charge analysis taken account within the Wigner-Seitz atomic radius (Table I), the charges of 0.02 and 0.21$e^{-}$ are depleted from the Zn and O atoms, respectively.

As the stability of the layered ZnO can be enhanced by an epitaxial strain \cite{wu11},
we next explore the impact of biaxial compressive and tensile strains ($\varepsilon$) on magnetism of the Zn-vacancy defected ZnO monolayer,
where $\varepsilon$ $=$ (\emph{a$_{\|}$} -- \emph{a$_0$})/\emph{a}$_0$$\times$100\%;
$a_0$ is the equilibrium lattice (3.296 \AA) of the Zn-vacancy defected ZnO and $a_{\|}$ is a variable.
We here would like to note that either in-plane lattice expansion and compression up to $\sim$3\% of graphitic ZnO layers
can be practically achieved by controlling the number of atomic layers \cite{lee16}.
The magnetic moments of Zn (blue symbol) and O (black symbol) atoms are shown in Fig. 3(a) for different strains, ranging from a compressive strain of --3\% to a tensile strain of $+$3\%.
The Zn moment decreases as strain increases (compressive $\rightarrow$ tensile), whereas it increases for the O case, which is a reflection of Vergard's law \cite{vegard21}.

\begin{table} [t]
\caption{The atomic contributions of Zn and O species to the total MAE (meV) of a single-layer ZnO with a Zn-vacancy for different strains $\varepsilon$ (\%),
where the corresponding in-plane lattice constants \emph{a} (\AA) are also listed.}
\begin{ruledtabular}
\begin{tabular}{cccccc}
$\varepsilon$ & \emph{a} & MAE(Zn) & MAE(O) & MAE(Total) \\
\hline
--2 & 3.230 & $+$0.05 & --0.01 & $+$1.06  \\
0 & 3.296 & --0.05 & --0.01 & --0.50  \\
2 & 3.361 & --0.03 & --0.01 & --0.29  \\
\end{tabular}
\end{ruledtabular}
\end{table}

Figure 3(b) shows the calculated MAE of the Zn-vacancy defected ZnO versus strain.
The result of the negative MAE (--0.5 meV) at zero strain indicates that the induced magnetism prefers an in-plane spin orientation parallel to the lateral lattice.
Both the compressive and tensile strains enhance MAE.
More interestingly, the sign of MAE is reversed into the positive and reaches up to $+$1 meV at --2\% strain.
This indicates that the magnetization direction can reorient
from in-plane to perpendicular orientation (PMA) upon an application of compressive strain.
Such a PMA remains at strain up to the largest strain of --3\% considered in our study.
On the other hand, the MAE tends to disappear (or reorient)
at the larger tensile strain beyond $+$3\% as the Zn moment reaches zero.

\begin{figure} [t]
  \centering
 \includegraphics[width=1\columnwidth]{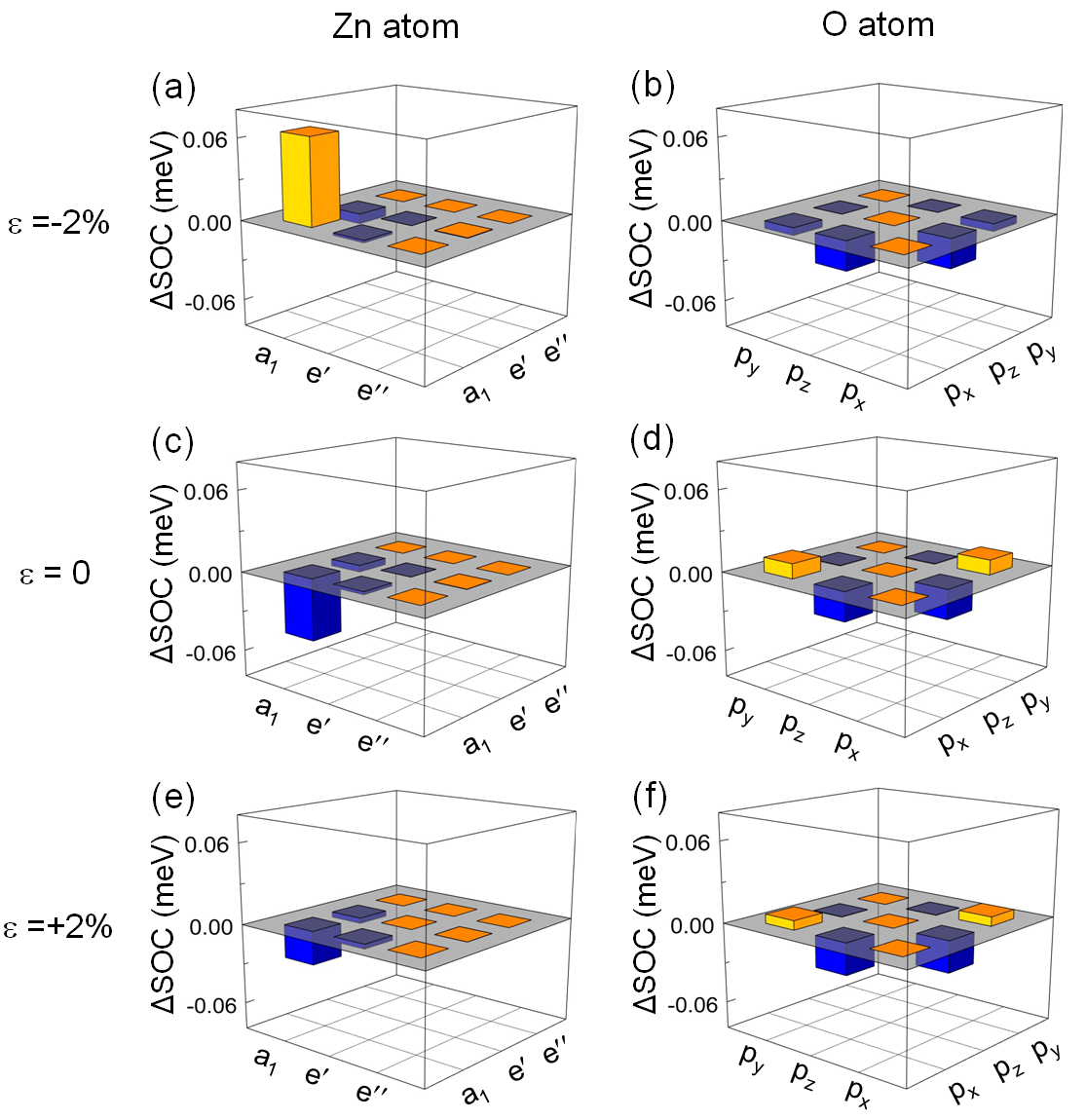}
  \centering
  \caption{Difference of the orbital-projected SOC energies, $\Delta E_{\textrm{soc}}$, between in- and out-of-plane magnetization orientation of Zn and O atoms in a single-layer ZnO with a Zn-vacancy for the (a) and (b) --2\% and (c) and (d) zero and (e) and (f) $+$2\% strain, respectively. The yellow and blue bars represent the positive and negative $\Delta E_{\textrm{soc}}$, respectively.}
\end{figure}

The contributions of different atomic species (Zn and O) to the total MAE are shown in Table II for the selected strains of $\varepsilon$ $=$ --2, 0, and $+$2\%,
from the difference of SOC energies between in- and out-of-plane magnetization, $\Delta E_{\textrm{soc}}$ $=$ $E_\textrm{soc}^\|$ -- $E_\textrm{soc}^\bot$.
Here, $E_{\textrm{soc}} = <\frac{\hbar^2}{2m^2c^2}\frac{1}{r}\frac{dV}{dr} \textbf{\emph{L}}\cdot\textbf{\emph{S}}>$,
where \emph{V}(\emph{r}) is the spherical part of the effective potential within the PAW sphere, and \textbf{\emph{L}} and \textbf{\emph{S}} are
orbital and spin operators, respectively.
The expectation value of $E_{\textrm{soc}}$ is twice the actual value of the total energy correction to the second-order in SOC, i.e., MAE $\approx$ 1/2$\Delta E_{\textrm{soc}}$ \cite{antropov14,zhang17}.
Our test calculations indicate that the second-order perturbation theory is a reasonable approximation as the total MAE overall agree within a few percent accuracy
with those obtained from the atom and orbital projected calculations.
The other 50\% of the SOC energy translates into the crystal-field energy and the formation of the unquenched orbital moment \cite{skomski11}.

\begin{figure} [t]
  \centering
 \includegraphics[width=1\columnwidth]{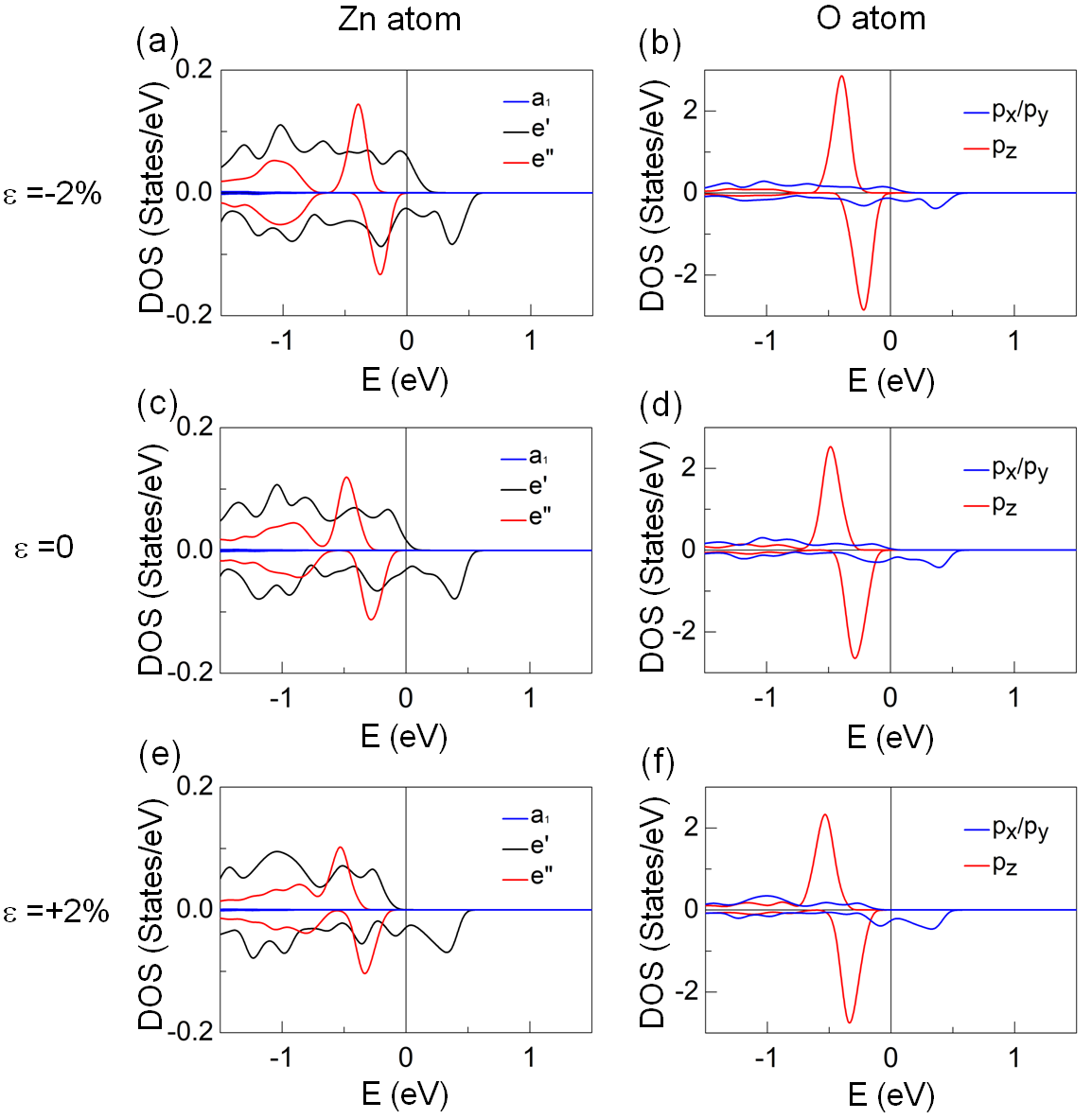}
  \caption {Spin- and orbital-decomposed PDOS of Zn and O atoms of the defective ZnO monolayer with a Zn-vacancy for the (a) and (b) $\varepsilon=-2\%$ and (c) and (d) $\varepsilon=0$ and (e) and (f) $\varepsilon=+2\%$, respectively. The singlet $a_1$ ($d_{z^2}$) and doublet $e'$ ($d_{xy/x^2−y^2}$) and $e''$ ($d_{xz/yz}$) orbital states of Zn atom are shown in blue, black, and red lines, respectively. The $p_{x,y}$ and $p_{z}$ orbital states of O atom are shown in blue and red lines, respectively. The Fermi level is set to zero in energy.}
\end{figure}

As seen in Table II, the $\Delta E_{\textrm{soc}}$ of Zn atom exhibits a trend similar to the changes in sign of the total MAE under strain, whereas the contribution of O atom is rather small and strain-independent.
Thus, the dominant contribution to the MAE must come from the Zn atom, where the larger SOC effect of \emph{d} orbitals ($l = 2$) than \emph{p} orbitals ($l=1$) as a physics origin of anisotropic phenomena is well manifested.
The validity of this argument is further demonstrated in connection with the Bruno theory
$ \textrm{ MAE} = \frac{\xi}{4\mu_B}\Delta \mu_{\textrm{o}}$ \cite{Bruno1989},
where $\Delta \mu_{\textrm{o}}$ $=$ $\mu_\textrm{o}^\bot - \mu_\textrm{o}^\|$ is the orbital magnetic anisotropy (OMA) and $\xi$ is the SOC constant.
The calculated $\Delta \mu_{\textrm{o}}$ of the Zn and O atoms are --0.6$\times$ and 0.8$\times$10$^{-2}$ $\mu_{B}$ at zero strain, respectively.

To get more insights on the strain-induced magnetization reversal, we further decompose the $\Delta E_{\textrm{soc}}$
into the \emph{d}-orbital matrix of the Zn atom for --2, 0, and $+$2\% strains in Fig. 4.
For comparison, we also plot the same for the \emph{p}-orbital matrix of O atom (right panels in Fig. 4).
The O \emph{p}-orbital contributions of MAE are relatively small and insensitive,
compared with the Zn \emph{d}-orbital contributions,
to the sensitivity of MAE under strain effect, as discussed in atom-decomposed MAE shown in Table II.
We thus focus on the \emph{d}-orbital states of the Zn atom for the discussion of the electronic origin for magnetization reversal.

\begin{figure} [t]
  \centering
 \includegraphics[width=1\columnwidth]{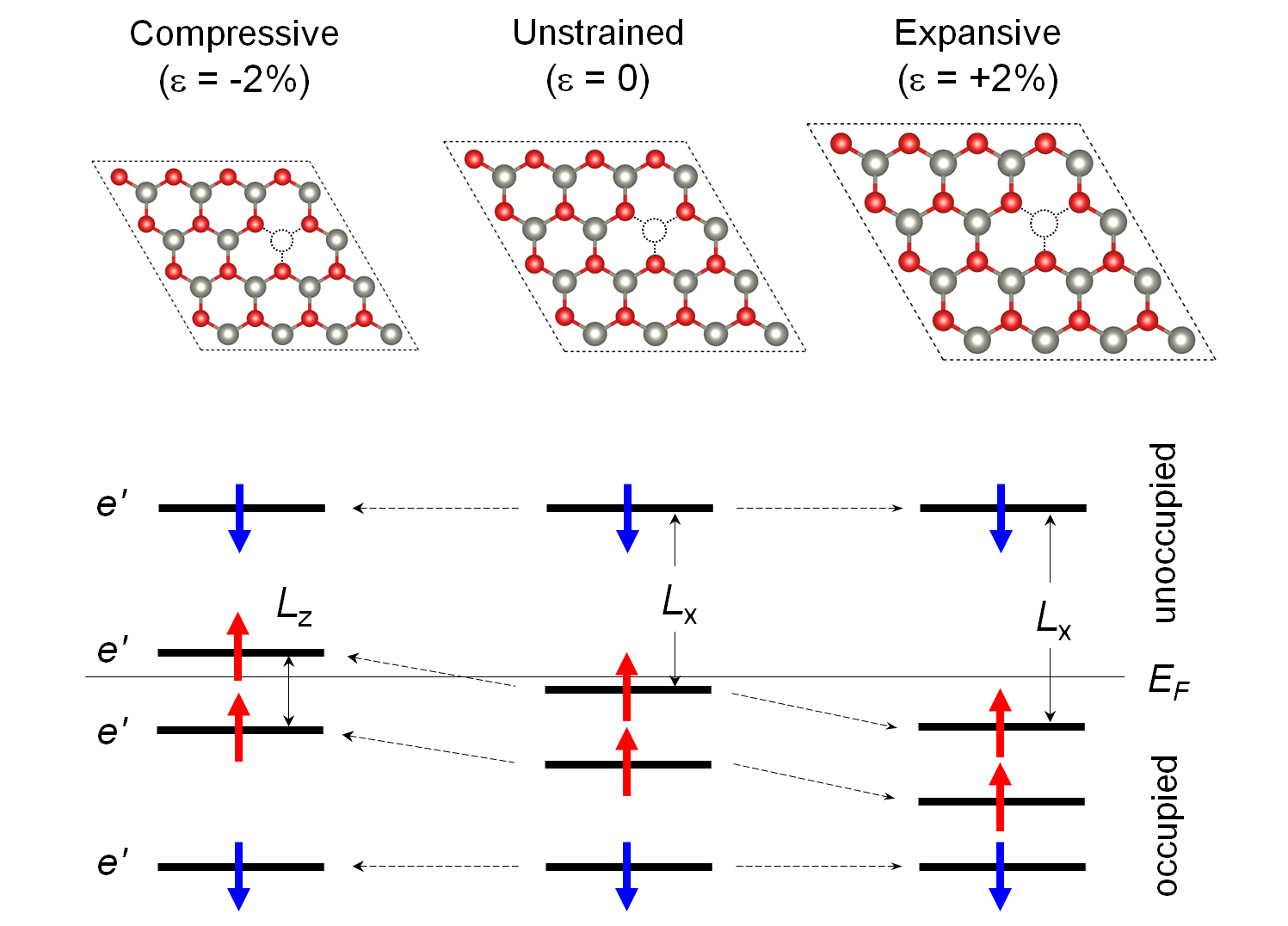}
  \caption {Energy levels of Zn $e'$ orbital states under the compressive ($\varepsilon$ $=$ $-$2\%) and expansive ($\varepsilon$ $=$ $+$2\%) strain.
  The upward red and downward blue arrows denote the majority-spin and
minority-spin states, respectively. The thin horizontal line indicates the Fermi level ($E_F$). The minority-spin $e'$ states are not much perturbed under strain.
The spin-orbit coupling pairs near the Fermi level between the occupied and unoccupied states are emphasized, where the spin up-up and up-down couplings
provide the positive (by $\hat{L}_{z}$) and negative contribution (by $\hat{L}_{x}$) to the MAE, respectively.
The atomic symbols at the top (crystallographic image) are the same as used in Fig. 1.}
\end{figure}

\begin{figure} [b]
  \centering
 \includegraphics[width=1\columnwidth]{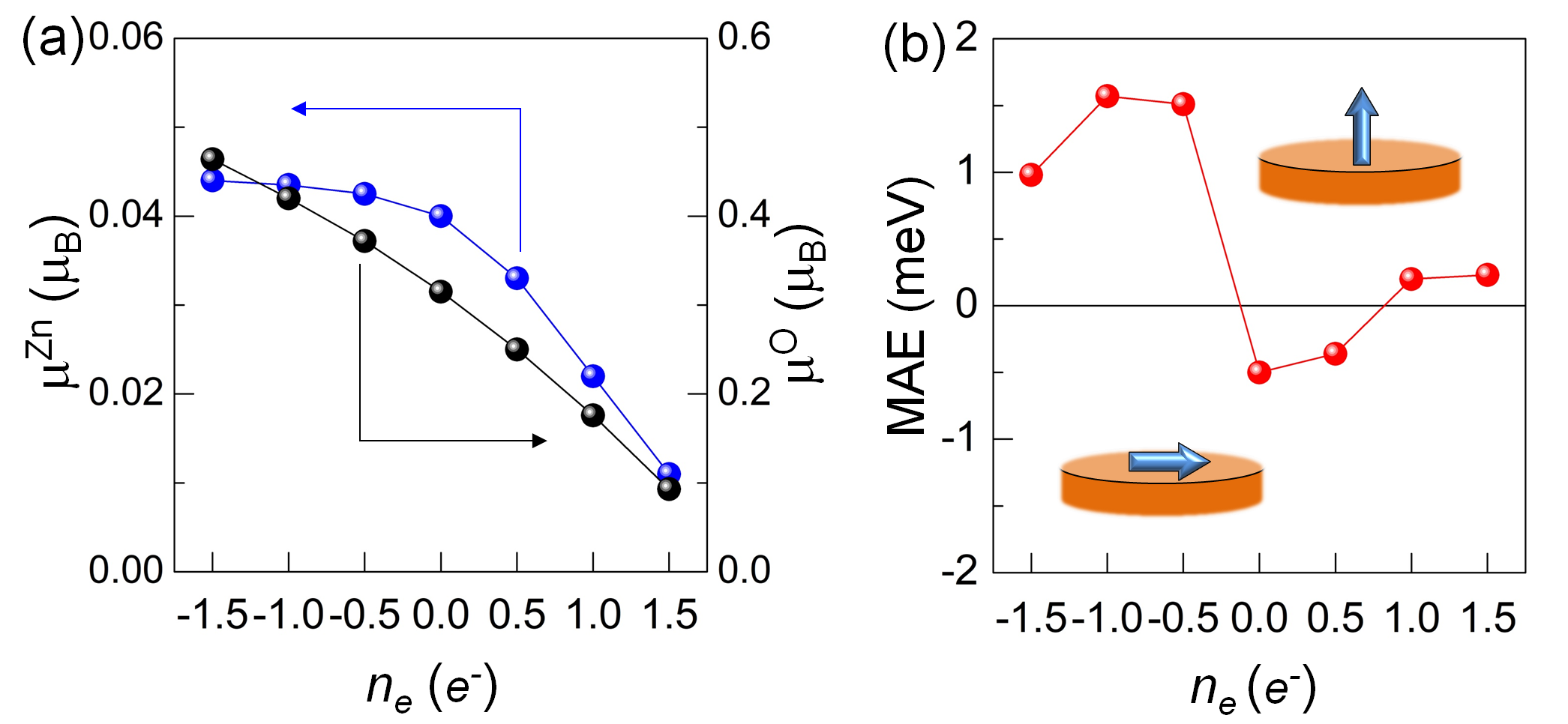}
  \caption {(a) Magnetic moments $\mu$ of Zn (blue) and O (black) atoms and (b) MAE as a function of the number of extra valence electrons $n_e$ ($e^{-}$) in a 4$\times$4 unit cell of ZnO monolayer with a Zn-vacancy.
  The negative and positive values in $n_e$ correspond to the charge deficit and addition, respectively.
    The insets in (b) show the schematics for magnetization switching from in-plane (bottom in negative MAE) to out-of-plane magnetization (top in positive MAE) at $n_e=-0.5e^{-}$ and $n_e=+1e^{-}$.}
\end{figure}

In the second-order perturbation theory, the MAE is determined by the SOC between occupied and unoccupied \emph{d}-orbital states \cite{wang93};
$
  \textrm{MAE}^{\sigma\sigma'} = \xi^2 \sum_{o,u}
  \frac{|\langle \Psi_{o, \sigma} | \hat{L}_{z} | \Psi_{u, \sigma'} \rangle |^2 - {|\langle \Psi_{o, \sigma} | \hat{L}_{x} | \Psi_{u, \sigma'} \rangle |^2}}
  { E_{u, \sigma'} - E_{o, \sigma} },
$
where $\Psi_{o}$ ($\Psi_{u}$) and $E_{o}$ ($E_{u}$) are the eigenstates and eigenvalues of occupied (unoccupied) states (band index \emph{n} and wave vector \textbf{k} are omitted for simplicity) for each spin state, $\sigma, \sigma'$ $=$ $\uparrow, \downarrow $, respectively, and $\hat{L}_{x(z)}$ is the \emph{x} (\emph{z}) component of the orbital angular momentum operator.
For $\sigma\sigma'$ $=$ $\uparrow \uparrow$ or $\downarrow \downarrow$, the positive (negative) contribution to MAE is determined by the
SOC interaction between the occupied and unoccupied states with the same (different by one) magnetic quantum number (\emph{m}) through the $\hat{L}_{z}$ ($\hat{L}_{x}$) operator.
Relative contributions of the nonzero $\hat{L}_{z}$ and $\hat{L}_{x}$ matrix elements are $ \langle xz|\hat{L}_{z}|yz\rangle $ $=$ 1, $ \langle xy|\hat{L}_{z}|x^2-y^2\rangle $ $=$ 2,
$ \langle xz,yz|\hat{L}_{x}|z^2\rangle $ $=$ $\sqrt{3}$, $ \langle xz,yz|\hat{L}_{x}|xy\rangle $ $=$ 1, and $ \langle xz,yz|\hat{L}_{x}|x^2-y^2\rangle $ $=$ 1 \cite{wang93}.
For $\sigma\sigma'$ $=$ $\uparrow \downarrow $, MAE has the opposite sign, so the positive (negative) contribution comes from the $\hat{L}_{x}$ ($\hat{L}_{z}$) coupling.

\begin{figure} [t]
  \centering
 \includegraphics[width=1\columnwidth]{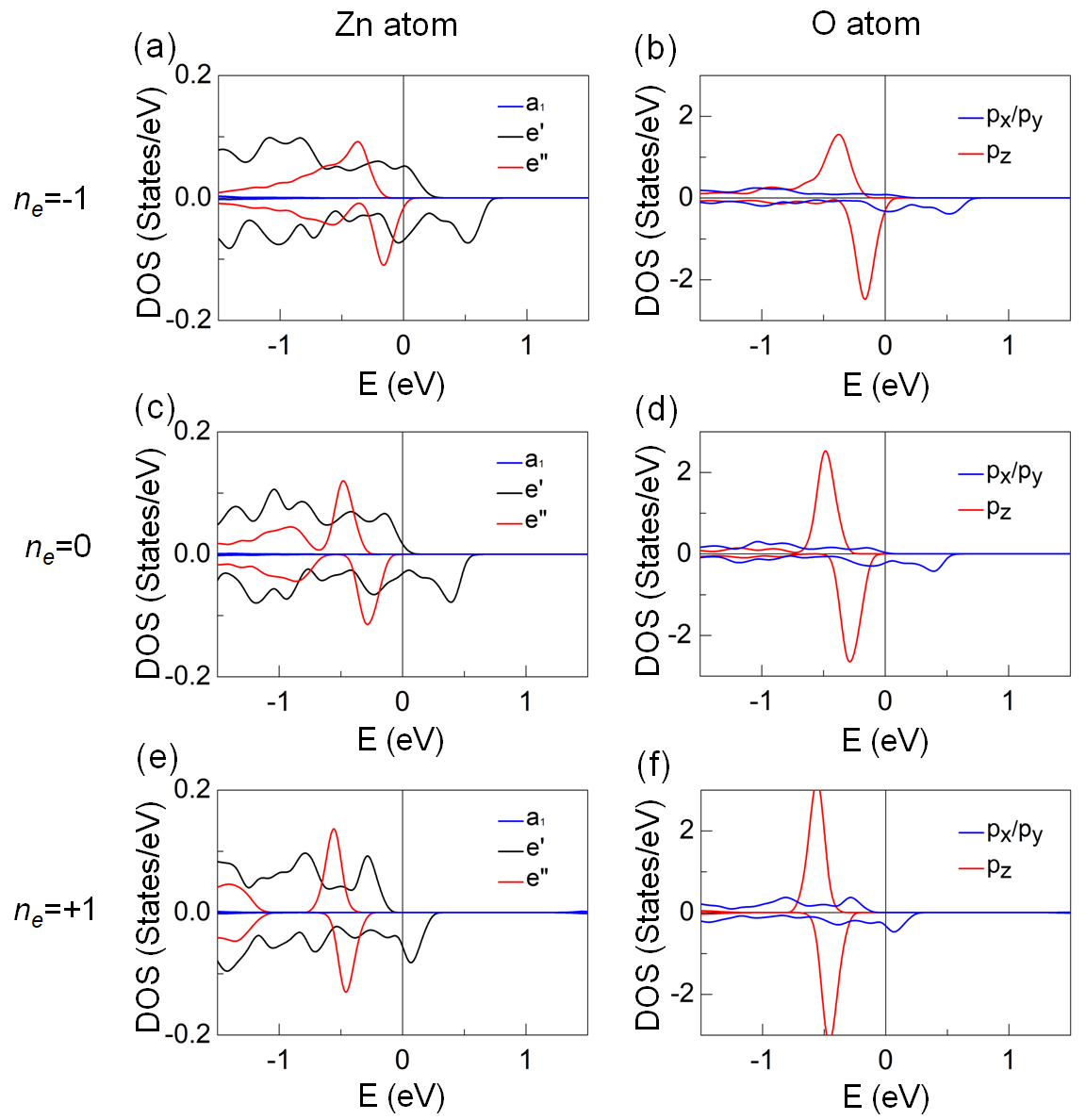}
  \caption {Spin- and orbital-decomposed PDOS of Zn and O of the defective ZnO monolayer with a Zn-vacancy for the (a) and (b) $n_e = -1e^{-}$ and (c) and (d) $n_e = 0$ and (e) and (f) $n_e = +1e^{-}$, respectively. The singlet $a_1$ ($d_{z^2}$) and doublet $e'$ ($d_{xy/x^2−y^2}$) and $e''$ ($d_{xz/yz}$) orbital states of Zn atom are shown in blue, black, and red lines, respectively. The $p_{x,y}$ and $p_{z}$ orbital states of O atom are shown in blue and red lines, respectively. The Fermi level is set to zero in energy.}
\end{figure}

Figures 5(a)--5(f) show the \emph{d} (\emph{p})-orbital PDOS of the Zn (O) atom at $\varepsilon = -2$, 0, and $+$2\%, respectively.
In Fig. 6, we also illustrate simple energy level diagram for the changes of Zn $e'$ peak states
near the Fermi level under strain.
The electronic structure analyses along with the \emph{d}-orbital decomposed MAE indicate that the underlying origin of the negative MAE at zero strain
is the spin-orbit coupling between the majority-spin occupied and the minority-spin unoccupied $e'$ states through the in-plane orbital angular momentum operator,
$ \langle xy _ \uparrow| \hat {L}_{x}|x^2-y^2 _ \downarrow \rangle $ (or $ \langle x^2-y^2  _ \uparrow| \hat {L}_{x}|xy_ \downarrow \rangle $).
This negative contribution weakens at $+$2\% strain because of the downward shift of the majority-spin filled $e'$ states away from the Fermi level.
For the compressive strain,
the positive MAE arises primarily from the same spin-channel ($\uparrow \uparrow$) matrix elements by
$ \langle xy _ \uparrow| \hat {L}_{z}|x^2-y^2 _ \uparrow \rangle $
and $ \langle x^2-y^2 _ \uparrow| \hat {L}_{z}| xy_ \uparrow \rangle $, as the majority-spin degenerate $e'$ states become partially unoccupied.
Such a modification of the energy landscape of the Zn $e'$ electronic state
around the Fermi level is the result of the changes in the strength of orbital hybridization with the in-plane O $p_{x/y}$ states under strain.

The engineering of MAE by Fermi level shifts further suggests to explore crucial effects of the external gating and chemical doping on magnetization reversal.
To anticipate this phenomenon, we have performed electrostatic doping calculations
by changing the number of valence electrons ($n_e$) in the whole system (4$\times$4 unit cell of ZnO monolayer)
from $n_e=-1.5$ to $+1.5e^{-}$.
This approach reflects to the charge carrier density that is confined over the metallic surface rather than exceeding to a vacuum region, which is analogues to the electric field effect by the positive and negative gating as well as chemical doping with the neighboring elements (e.g., Cu or Ga for Zn, and N or F for O site) \cite{zhang17}.

The calculated magnetic moments and MAE as a function of $n_e$ are shown in Figs. 7(a) and 7(b), respectively.
It is expected that the charge deficit enhances the magnetic moments of both Zn and O atoms
whereas the charge excess reduces because of the electron/hole doping.
Nevertheless, as shown in Fig. 7(b), both the deficit and excess valence charges promptly enhance the MAE
and change its sign from in-plane to perpendicular magnetization at $n_e=-0.5$ and $=+1e^-$, respectively.
These results are of considerable interest in the area of the voltage-controlled magnetic anisotropy (VCMA) and magnetoelectric phenomena.
In the present study, the VCMA can be determined by the curve slope of the electric-field or $n_e$ dependence of MAE.
The nonlinear parabola-like shape behavior of the VCMA within a small range of $n_e$ ($-1<n_e<+1$)
is quite typical in most magnetoelectric materials, as reported in previous calculations \cite{zhang17,odkhuu15,odkhuu16} and experiments \cite{raj13,noz13,noz16}.

In Figs. 8(a)--8(f), we plot the \emph{d} (\emph{p})-orbital PDOS of the Zn (O) atom for $n_e=-1$, 0, and $+1e^{-}$, respectively.
The entire band structures of both Zn and O atoms shift toward the low-energy region upon the presence of external electrons from $-1$ to $+1e^{-}$ due to the band filling effect.
Since the majority-spin Zn $e'$ peak exists at the Fermi level, the similar argument discussed for the --2\% strained ZnO applies
to the origin of PMA at $n_e=-1e^{-}$.
On the other hand, for $n_e=+1e^{-}$
the peak position of the minority-spin empty $e'$ bands appears just above the Fermi level while the majority-spin states are fully occupied, as shown in Fig. 8(e).
Thus the SOC pair between the minority-spin filled and empty $e'$ states around the Fermi level
leads to the positive MAE through
$ \langle xy _ \downarrow| \hat {L}_{z}|x^2-y^2 _ \downarrow \rangle $
and $ \langle x^2-y^2 _ \downarrow| \hat {L}_{z}| xy_ \downarrow \rangle $.

\section{\textbf{CONCLUSION}}
In summary, we have performed the first-principles electronic structure calculations to reveal the critical effects of strain and external gating on magnetism and magnetic anisotropy of a single-layer ZnO with a Zn-vacancy.
In contrast to the pristine and defective ZnO with an O-vacancy, the presence of a Zn-vacancy
induces significant magnetic moments to its first neighboring O and Zn atoms due to the charge deficit.
More remarkably, we show that a compressive biaxial strain of only $\sim$1--2\% or either negative and positive gate voltages
give rise to magnetization reorientation from in- to out-of-plane magnetization.
The further analyses provide the underlying mechanism for magnetization reversal,
which is the result of the strain- and electric field-induced changes in spin-orbit coupled Zn \emph{d}-orbital states hybridized with O \emph{p} states.
These predictions may inspire further experimental and theoretical explorations
of exploiting epitaxial biaxial strain and electric field to manipulate magnetism and magnetization direction for two-dimensional spintronic applications.

This work was supported by Postdoctoral Research Program (2017) through Incheon National University.

\end{document}